\documentclass[%
 reprint,
 amsmath,amssymb,
 aps,
]{revtex4-2}

\usepackage{graphicx}
\usepackage{dcolumn}
\usepackage{bm}

\begin{document}

\preprint{APS/123-QED}

\title{Mass and infinite dimensional geometry}

\author{Puskar Mondal}
\email{puskar\_mondal@fas.harvard.edu}
\affiliation{%
 Centre of Mathematical Sciences and Applications, Harvard University, 20 Garden Street, Cambridge, USA\\
 Department of Mathematics, Harvard University, 1 Oxford Street, Cambridge, USA 
}%





\begin{abstract}
\noindent I unravel an elegant geometric meaning of the mass of the lowest energy excited state of a renormalizable quantized field theory by studying the weighted geometry of the classical configuration space of the theory. A suitably defined regularized Bakry-Emery Ricci curvature of these infinite dimensional spaces controls the spectra of the corresponding quantum Hamiltonians. The Ricci curvature part of the full Bakry-Emery Ricci curvature appears to be purely quantum in nature. This geometric contribution to the spectra in the context of quantum field theory has not been studied previously to my knowledge. Assuming the existence of rigorous quantization, I present a few problems starting from massive free particles to the non-abelian Yang-Mills theory. A remarkable property is observed in the large $N$ Yang-Mills theory, where a non-trivial mass gap is preserved. This occurs due to the fact that the regularized Bakry-Emery Ricci curvature that is responsible for the gap of the configuration space scales as $g^{2}_{YM}N=\lambda$ ('t Hooft coupling) that remains invariant. 
\end{abstract}

\maketitle

\noindent Quantum field theory has been extremely successful in describing the interactions of elementary particles. While a rigorous mathematical treatment of $3+1$ dimensional interacting quantum field theories remains intractable, several interesting physical consequences have been discovered and experimentally verified. The most spectacular advances are in the sector of gauge theory which is the building block of the standard model of particle physics. One of the most important physical results that are available in the framework of gauge theory is the asymptotic freedom of non-abelian gauge theory \cite{gross1973ultraviolet}. This discovery marked a sharp distinction between the non-abelian gauge theory and its abelian counterpart. Another problem that remains elusive to this day is the existence of a positive mass gap of quantum Yang-Mills theory formulated with a non-abelian compact gauge group \cite{jaffe2006quantum}. Lattice gauge theory calculations indicated the existence of such gap supporting the short-range property of the strong force or essentially the non-existence of free mass-less gluons \cite{lucini2010glueball, athenodorou2020glueball}. In lower spacetime dimensions (e.g., $2+1$ dimensions), calculations supporting a gap in the spectra of Hamiltonian are performed in \cite{karabali1996gauge,karabali1996gauge,karabali1997gauge,karabali1998planar, karabali1998vacuum}. In these lower dimensional examples, the kinetic part of the quantum Hamiltonian played an extremely important role. Since the kinetic part is nothing but a type of functional Laplacian on the \textit{true} configuration space, it becomes natural to investigate the geometry of the configuration space and attempt to obtain a gap estimate in terms of curvature (of the \text{true} configuration space or orbit space). A possible complication is that a true gap estimate should take into account the potential part as well. As it turns out that the geometric analysis in the presence of a potential can be cast into a problem of analysis on a suitably defined `weighted' manifold. In the broader context of bosonic quantum field theory, I address: can the mass of the least energy excited state in a bosonic quantum field theory have a geometric interpretation? In addition, there have been studies based on more direct approach such as solving the Schwinger-Dyson equations \cite{frasca1,frasca2} in both $2+1$ and $3+1$ dimensional cases. 

I adopt a Hamiltonian formalism of the quantum field theory on $(\mathbb{R}^{1,n},\eta)$, $\eta$ being the standard Minkowski metric. The Hamiltonian perspective is well established and proven to be equivalent to the path integral formulation for renormalizable field theories (see \cite{jackiw1988analysis,symanzik}). Even though Lorentz covariant techniques such as the path integral method are well suited for perturbative calculations contrary to the non-covariant Hamiltonian picture, the latter is believed to be conducive to non-perturbative aspects of quantum field theory. One of the most interesting perspectives is to give the classical field $\varphi(x)$ a particle interpretation. After all, one can view a field $\varphi(x)$ as a collection of mechanical variables $q_{i}$ ($i=1,2,3,......., N$) for $N$ degrees of freedom in the limit that $N$ becomes uncountably infinite. Let us consider the space of fields $\Phi:=\{\varphi|\varphi~\text{is a section of an appropriate bundle over}~\mathbb{R}^{1,3}$ $ ~\text{and}~\varphi~\text{lies~in~a~suitable~function~space}\}$ (for gauge theories one identifies gauge equivalent fields to yield a reduced space $\widehat{\Phi}$). In $n+1$ formulation, the dynamics of a classical field $\varphi$ in the configuration space $\Phi$ can be thought of as a continuous curve $I\subset \mathbb{R}\to \Phi,~t\mapsto \varphi(t)$ with prescribed initial position $\varphi_{0}\in \Phi$ and momentum $\dot{\varphi}_{0}\in T_{\varphi_{0}}\Phi$. As is well known such an interpretation breaks down at the quantum level even in the finite dimensional setting. The quantization yields wave functionals on the configuration space. Let us consider a field theory with action $S=\int_{\mathbb{R}^{1,3}}\mathcal{L} d^{4}x$ associated to a real classical field $\varphi$ (here $\varphi$ can be a massive scalar field or a gauge field taking its values in suitable vector bundles, complex fields can also be handled as we shall see while addressing the electroweak sector). The associated Hamiltonian $H$ is a conserved entity on $\mathbb{R}^{1,3}$ due to the presence of a time-like Killing field. The canonical quantization amounts to promoting the field $\varphi$ and its functionals to operators on a separable Hilbert space $\mathcal{H}$. The section $\varphi$ is then the eigenstate of the associated operator $\widehat{\varphi}$ i.e., $\widehat{\varphi}|\varphi\rangle=\varphi(x)|\varphi\rangle$. The next step is to define an equal time commutation condition between the field operator $\widehat{\varphi}$ and its conjugate momentum operator $\widehat{\pi}$, solve the associated functional Schr\"odinger's equation to obtain wave-functionals $\Psi[\varphi]:=\langle\varphi|\Psi\rangle$, and impose functional constraints (for gauge theory) on the wave functionals. All the classical conserved entities associated with isometries of $(\mathbb{R}^{1,3},\eta)$ become operators on $\mathcal{H}$. The energy that is the time component of the energy-momentum $4-$vector is naturally defined in the quantum theory as the eigenvalue of the operator (functional) Hamiltonian. In the process, one is required to regularize and normal order the associated functional Hamiltonian. Strictly speaking, however, the excited states are not eigenstates of the Hamiltonian.  

In this article, I will use the following dimensions for the entities involved. I set the light speed $c$ equal to $1$. With this convention, I have $[t]=[x]=L$, $L$ being the length dimension. The classical action has the dimension of $\hbar$. Therefore, the dimensions of the fields and coupling constants follow accordingly. Let us now endow the space of fields $\Phi$ (or the true configuration space $\widehat{\Phi}$) by a Riemannian metric that naturally arises from the kinetic part of the Lagrangian. The Lagrangian may be written as follows 
\begin{eqnarray}
\mathcal{L}=\frac{1}{2}\int_{\mathbb{R}^{3}\times \mathbb{R}^{3}}\mathcal{G}_{\varphi(x)\varphi(y)}\partial_{t}\varphi(x)\partial_{t}\varphi(y)-\text{Potential}
\end{eqnarray}
where $\mathcal{G}:T_{\varphi}\Phi\times T_{\varphi}\Phi\to \mathbb{R}_{+}$ is a Riemannian metric on $\Phi$ induced by the kinetic energy. Through a Legendre transformation, one may compute the inverse metric $(\mathcal{G}^{-1})^{\varphi(x)\varphi(y)}$ and write the action in terms of the conjugate momentum $\pi(x):=\frac{\delta \mathcal{L}}{\delta \partial_{t}\phi(x)}\in T^{*}_{\varphi}\Phi$ 
\begin{eqnarray}
\label{eq:legendre}
\mathcal{L}=\frac{1}{2}\int_{\mathbb{R}^{3}\times \mathbb{R}^{3}}(\mathcal{G}^{-1})^{\varphi(x)\varphi(y)}\pi(x)\pi(y)-\text{Potential}.
\end{eqnarray}
In the canonical quantization of the theory, one promotes $\varphi$ and $\pi$ to operators $\widehat{\varphi}$ and $\widehat{\pi}$ on the associated Hilbert space and imposes the equal time commutation condition 
\begin{eqnarray}
[\widehat{\varphi}(t,x),\widehat{\pi}(t,y)]=i\delta_{*}(x,y).
\end{eqnarray}
Here $\delta_{*}(x,y)$ may not always be the usual Dirac distribution. Especially in gauge theories, this would depend on the choice of gauge (e.g., this will be a transverse Dirac distribution for quantized Maxwell theory in Coulomb gauge). 
The formal Hamiltonian operator of renormalizable quantized theory may now be written using the relation (\ref{eq:legendre}) as follows 
\begin{eqnarray}
\widehat{H}:=-\frac{\hbar^{2}}{2}\int_{\mathbb{R}^{3}\times \mathbb{R}^{3}}(\mathcal{G}^{-1})^{\varphi(x)\varphi(y)}\frac{\mathfrak{D}}{\mathfrak{D}\varphi(x)}\frac{\mathfrak{D}}{\mathfrak{D}\varphi(y)}\\\nonumber 
+\text{Potential},
\end{eqnarray}
where $\frac{\mathfrak{D}}{\mathfrak{D}\varphi}$ is the compatible connection induced by the metric $\mathcal{G}$ on the tangent bundle $T\Phi$ (there are certain restrictions for a metric compatible connection to existing on an infinite dimensional manifold. See \cite{freed1989basic} for detail. I choose an appropriate regularity of $\varphi$ so that the existing criteria are satisfied e.g., I assume the field is an element of the Schwartz space). This Hamiltonian, however, is formal in the sense that the covariant functional Laplacian $\int_{x,y}(\mathcal{G}^{-1})^{\varphi(x)\varphi(y)}\frac{\mathfrak{D}}{\mathfrak{D}\varphi(x)}\frac{\mathfrak{D}}{\mathfrak{D}\varphi(y)}$ is ill-defined even on smooth functionals. Therefore a certain regularization is necessary to make sense of this as an infinite dimensional elliptic operator on the configuration space $\Phi$. Notice that in field theories $(\mathcal{G}^{-1})^{\varphi(x)\varphi(y)}$ has a structure of the type $(\mathcal{G}^{-1})^{\varphi(x)\varphi(y)}=\delta(x-y)+\cdot\cdot\cdot\cdot$ and therefore the leading order term in the covariant functional Laplacian is the ordinary flat space functional Laplacian $\int_{\mathbb{R}^{3}}\frac{\delta^{2}}{\delta\varphi(x)\delta\varphi(x)}$ that is ill-defined. Therefore a natural choice of regularization would be to replace the usual Dirac's distribution with a point split distribution \cite{hatfield1984first,karabali1998planar} $\delta_{\chi}(x,y)$ that weakly recovers $\delta(x,y)$ in the limit $\chi\to\infty$ i.e., 
\begin{eqnarray}
\int\delta_{\chi}(x,y)\xi(y)\to\xi(x)~\text{as}~ \chi\to\infty.
\end{eqnarray}
For example, on $\mathbb{R}^{3}$, in standard coordinates, one may choose $\delta_{\chi}(x,y)$ as 
\begin{eqnarray}
\delta_{\chi}(x,y)=\frac{\chi^{3}}{\pi^{3}}e^{-\sum_{i=1}^{3}(x^{i}-y^{i})^{2}\chi^{2}}.
\end{eqnarray}
The second issue that arises is the ordering of $\widehat{H}$ so that the ground state has zero energy. However, in the current context where we are primarily interested in obtaining the energy required to excite the least mass state i.e., the difference between the energies of the ground state and the first excited state, the issue of normal ordering does not cause problems, and therefore will not be addressed here. Now, if I assume the existence of a rigorous quantum theory, it yields a normalizable ground state (one of the basic axioms of quantum field theory) $\Psi[\varphi]:=\langle \varphi|0\rangle$ that may be represented as 
\begin{eqnarray}
\label{eq:groundstate}
\Psi[\varphi]=N_{\hbar}e^{-\frac{S[\varphi]}{\hbar}},
\end{eqnarray}
where $S[\varphi]$ is a positive functional that grows sufficiently fast at infinity on the configuration space $\Phi$ such that $\Psi[\varphi]$ is normalizable i.e., 
\begin{eqnarray}
|N_{\hbar}|^{2}\int_{\Phi}e^{-\frac{2S[\varphi]}{\hbar}}\sqrt{\det(\mathcal{G})}=1.
\end{eqnarray}
I can use this normalization condition to yield a measure $|N_{\hbar}|^{2}e^{-\frac{2S[\varphi]}{\hbar}}\sqrt{\det(\mathcal{G})}$ on the configuration space $\Phi$ (for an interacting theory this measure is non-Gaussian). Once such a measure is obtained, I am allowed to perform the elliptic analysis on the weighted manifold or the metric measure space $(\Phi, \mathcal{G},|N_{\hbar}|^{2}e^{-\frac{2S[\varphi]}{\hbar}}\sqrt{\det(\mathcal{G})})$. The Hilbert space then takes the form of a $L^{2}$ space with respect to the aforementioned measure $L^{2}\left(\Phi,\mathcal{G},|N_{\hbar}|^{2}e^{-\frac{2S[\varphi]}{\hbar}}\sqrt{\det(\mathcal{G})}\right)$. With the point-split regularization of the functional Hamiltonian operator, I have the following theorem for its spectral gap or mass gap. To be more precise, I denote the mass gap (the energy difference between the ground state and the first excited state) by $\Delta E$. This theorem is proven elsewhere \cite{puskar2023geometric} and therefore we omit the proof here.\\    
\textbf{Theorem (gap)}
\textit{Under the assumption of the existence of a quantum field theory for the field $\varphi\in\Phi$ (scalar field or gauge field taking its value in a suitable bundle) and the following positivity of the regularized Bakry-Emery Ricci curvature of the space} $\Phi$
\begin{eqnarray}
\text{Ricci}^{B.E}(\alpha[\varphi],\alpha[\varphi])\nonumber \geq \Delta \int_{\mathbb{R}^{3}\times \mathbb{R}^{3}}\mathcal{G}^{\varphi(x)\varphi(y)}\alpha_{\varphi(x)}[\varphi]\alpha_{\varphi(y)}[\varphi],\\\nonumber 
\Delta>0,\\\nonumber 
\mathcal{R}icci^{B.E}(\alpha[\varphi],\alpha[\varphi]):=\\\nonumber 
\int\left((\mathcal{G}^{-1})^{\varphi(z)\varphi(y)}\mathcal{R}_{\varphi(z)}~^{\varphi(x^{''})}~_{\varphi(y)}~^{\varphi(x)}\alpha_{\varphi(x^{"})}[\varphi]\alpha_{\varphi(x)}[\varphi]\right.\\\nonumber 
\left.+\frac{2}{\hbar}(\mathcal{G}^{-1})^{\varphi(x)\varphi(x^{'})}(\mathcal{G}^{-1})^{\varphi(y)\varphi(y^{'})}\frac{\mathfrak{D}}{\mathfrak{D}\varphi(x)}\frac{\mathfrak{D}S}{\mathfrak{D}\varphi(y)}\right.\\\nonumber 
\left.\alpha_{\varphi(x^{'})}[\varphi]\alpha_{\varphi(y)}[\varphi]\right),
\end{eqnarray}
\textit{for a vector} $\alpha_{\varphi}[\varphi]\in T^{*}_{\varphi}\Phi$ \textit{and} $\mathcal{R}$ \textit{the Riemann curvature of $\Phi$, 
the regularized Hamiltonian operator verifies the following mass gap} 
\begin{eqnarray}
\Delta E\geq \frac{\hbar^{2}\Delta}{2}.
\end{eqnarray}
One important point to note here is that the gap $\Delta E$ depends on the regulator energy scale $\chi$. This is not surprising since the un-regulated Ricci curvature is divergent, and therefore, we are ultimately interested in the finite part of $\Delta$ through appropriate renormalization and taking the regularization limit i.e., $\chi\to\infty$ limit. One way to perform the renormalization is to subtract a $\chi$ dependent spacetime constant $C_{\chi_{0}}(\chi)$ ($\chi_{0}$ is an arbitrary subtraction scale that is much smaller than the regularization scale $\chi$) from the Hamiltonian that diverges as $\chi\to\infty$ in such a way that it cancels the divergence part of $\Delta$ in the spectra. At any finite value of $\chi$, the Hamiltonian is bounded from below. This formal manipulation, however, is to be understood in a concrete manner. This is reminiscent of the usual renormalization procedure performed in quantum field theory. I note that \cite{karabali2008robustness} addresses this issue of renormalization in the Hamiltonian picture (see their section 3).     
The main idea behind the proof is the commutation of covariant derivatives and integration by parts with respect to the measure $|N_{\hbar}|^{2}e^{-\frac{2S[\varphi]}{\hbar}}\sqrt{\det(\mathcal{G})}$. The proof is well known in the finite-dimensional framework (see \cite{li1980estimates}). 
The gap property of Hamiltonian is governed both by the kinetic contribution i.e., the trace geometry of the configuration space as well as the potential which manifests itself in terms of the functional $S[\varphi]$ appearing in the exponential of the ground state wave functional (\ref{eq:groundstate}). One can apply this theorem to the known cases and verify that it yields the exact result. I start with the massive free scalar field.\\
\section{Massive Free Scalar Field} \noindent Recall the free massive scalar field theory on the $3+1$ dimensional Minkowski space for which the exact ground state is available. The classical Lagrangian reads $\mathcal{L}[\xi]=-\frac{1}{2}\int_{\mathbb{R}^{3}}\eta^{\mu\nu}(\partial_{\mu}\xi\partial_{\nu}\xi+m^{2}\xi^{2}),~~\xi:\mathbb{R}^{1+3}\to \mathbb{R}~$ which may be explicitly written as 
\begin{eqnarray}
\mathcal{L}[\xi]=\int_{\mathbb{R}}\frac{1}{2}\int_{\mathbb{R}^{3}\times \mathbb{R}^{3}}\delta(x-y)\partial_{t}\xi(x)\partial_{t}\xi(y)\\\nonumber 
-\frac{1}{2}\int_{\mathbb{R}^{3}}(\eta^{ij}\partial_{i}\xi\partial_{j}\xi+m^{2}\xi^{2}),
\end{eqnarray}
$m$ denoting the mass. Therefore, in this case, I have $\varphi=\xi$. If we denote the configuration space by $\Phi$ as usual, then the kinetic term induces a flat Riemannian metric (in local coordinates $\xi$) on $\Phi$
\begin{eqnarray}
\mathcal{G}_{\xi(x)\xi(y)}=\delta(x-y).
\end{eqnarray}
The classical energy $E(k)$ has the following expression in terms of the mass  and $3-$momentum $k$,
$E(k)=\sqrt{k^{2}+m^{2}}$
i.e, $E(k)\geq m$. In the quantum version, the mass appears as a parameter of the irreducible representation of the Poincare group $SO(1,3)\ltimes \mathbb{R}^{1+3}$ the isometry group of the Minkowski space $\mathbb{R}^{1+3}$. In quantum field theory, this representation defines a one-particle Hilbert space $\mathcal{H}_{m}$ for a particular particle in the full spectrum of the particles. The full Hilbert space has the direct sum structure 
\begin{eqnarray}
\mathcal{H}=\mathbb{C}\oplus\left(\sum_{I}\oplus\mathcal{H}_{m_{I}}\right)\oplus m.p.s,
\end{eqnarray}
where $m.p.s$ denotes spaces of multi-particle states that are tensor products of one particle spaces. $\mathbb{C}$ corresponds to the ground state (vacuum) and has zero energy. Then there is a positive continuous spectrum starting from $\min_{I}(m_{I})=m$ and extending to infinity of the Hamiltonian (after normal ordered and regularized) of the theory 
\begin{eqnarray}
\widehat{H}:=-\int_{\mathbb{R}^{3}}\frac{\hbar^{2}}{2}\frac{\delta^{2}}{\delta\xi(x)\delta\xi(x)}\\\nonumber 
+\frac{1}{2}\int_{\mathbb{R}^{3}}(\eta^{ij}\partial_{i}\xi\partial_{j}\xi+m^{2}\xi^{2}).
\end{eqnarray}
According to our calculations, the spectral gap i.e., the least mass $m$ is supposed to be obtainable from the Bakry-Emery Ricci curvature associated with the infinite-dimensional weighted Riemannian manifold $(\Phi,\mathcal{G},|N_{\hbar}|^{2}e^{-2S[\xi]/\hbar})$, where $S[\xi]$ has the following explicit form \cite{hatfield2018quantum}
\begin{eqnarray}
S[\xi]=\frac{1}{2}\int_{k}\widehat{\xi}(k)\sqrt{k^{2}+m^{2}}\widehat{\xi}(-k)d^{3}k,
\end{eqnarray}
where $\widehat{\xi}$ is the Fourier transform (on $\mathbb{R}^{3}$) of the field $\xi(x)$.
Now since the metric $\mathcal{G}$ is flat, the Bakry-Emery curvature consists of the Hessian part of the $S$ functional only. An explicit calculation for the Bakry-Emery quadratic form in this particular case yields 
\begin{eqnarray}
\nonumber\text{Ricci}^{B.E}(\alpha[\xi],\alpha[\xi]):=\\\nonumber
\int\left((\mathcal{G}^{-1}_{\chi})^{\xi(z)\xi(y)}\mathcal{R}_{\xi(z)}~^{\xi(x^{''})}~_{\xi(y)}~^{\xi(x)}\alpha_{\xi(x^{"})}[\xi]\alpha_{\xi(x)}[\xi]\right.\\\nonumber 
\left.+\frac{2}{\hbar}(\mathcal{G}^{-1})^{\xi(x)\xi(x^{'})}(\mathcal{G}^{-1})^{\xi(y)\xi(y^{'})}\frac{\mathfrak{D}}{\mathfrak{D}\xi(x)}\frac{\mathfrak{D}S}{\mathfrak{D}\xi(y)}\right.\\\nonumber 
\left.\alpha_{\xi(x^{'})}[\xi]\alpha_{\xi(y)}[\xi]\right)
\geq\underbrace{0}_{\text{flatness of the configuration space}}\\\nonumber 
+\frac{2m}{\hbar}\int_{\mathbb{R}^{n}\times \mathbb{R}^{n}}(\mathcal{G}^{-1})^{\xi(x)\xi(y)}\alpha_{\xi(x)}[\xi]\alpha_{\xi(y)}[\xi],
\end{eqnarray}
or the energy gap $\Delta E \geq \hbar m$ from the gap theorem.
Notice that there is also a potential contribution in terms of the $3-$ momentum $k$ indicating a continuous spectrum starting from $m$ (i.e., actual estimate is $\Delta E\geq \hbar m+O(k^{2})$, the potential factor does not add a positive contribution). Therefore, the gap in the spectra of the Bakry-Emery curvature of the weighted configuration space $(\Phi,\mathcal{G}, N_{\hbar}e^{-2S[\xi]/\hbar})$ yields the mass gap or the lowest mass of the elementary particles. Since the configuration space is flat with respect to the induced metric (by the kinetic term), the mass gap is $m$ which is exactly what is expected. For a mass-less field, one would of course obtain a continuous spectrum starting from $0$.
Similarly, the case of a compact scalar (i.e., a scalar field theory on $\mathbb{T}^{3}\times \mathbb{R}$) can be handled easily through the Bakry-Emery Ricci curvature of the configuration space. In such case, the Hessian of the ground state would involve the Laplace-Beltrami operator of the compact manifold $\mathbb{T}^{3}$. Therefore, the spectra of the Hamiltonian have a strictly positive gap given by the gap in the spectra of the Laplace-Beltrami operator since in the compact case, the latter has a strictly non-zero lower bound depending on the diameter of the compact manifold in question.

\section{U(1) gauge field}\noindent  The $U(1)$ gauge field Lagrangian reads
\begin{eqnarray}
\mathcal{L}:=-\frac{1}{4}\int_{\mathbb{R}^{3}}F\wedge ~^{*}F,
\end{eqnarray}
where $F$ is the curvature $2-$form of the connection $A_{\mu}dx^{\mu}$ associated with a principle bundle on $\mathbb{R}^{1+3}$ with structure group $U(1)$ (pulled back along a suitable section of the bundle to be precise). The configuration space is not simply $\Phi:=\{\text{space of connections}~A~\text{lying in Schwartz function space}\}$ due to the invariance of the Lagrangian by a gauge transformation $A_{\mu}\mapsto A_{\mu}+\partial_{\mu}\lambda$ for an arbitrary smooth function $\lambda$. To construct the true configuration space of the theory, we must make the identification $A_{\mu}\sim A_{\mu}+\partial_{\mu}\lambda$ i.e., identify different points in $\Phi$. This amounts to taking a quotient of $\Phi$ by the group of gauge transformations (let us denote it by $\mathfrak{G}$). Therefore, the true configuration space turns out to be $\widehat{\Phi}:=\Phi/\mathfrak{G}$. The operation $\Phi\to \Phi/\mathfrak{G}$ at the level of the Lagrangian of the theory is performed by imposing the Gauss-Law constraint. More precisely, if we write down the connection $1-$form $A_{\mu}dx^{\mu}$ in the usual canonical coordinate as $(A_{0}, A_{i})_{i=1}^{3}$, then $A_{0}$ does not verify an evolution equation due to the fact that the corresponding momentum $\frac{\delta \mathcal{L}}{\delta(\partial_{t}A_{0})}$ vanishes. $A_{0}$ essentially acts as a Lagrange multiplier and generates the Gauss law constraint. The Gauss law constraint can in turn be used to express $A_{0}$ in terms of the dynamical variables $A_{i}$. More explicitly 
$\partial^{2}A_{0}=\partial_{t}\partial_{i}A_{i}$,
where $\partial^{2}$ is the spatial Laplacian on $\mathbb{R}^{3}$.
In order to study the geometric properties of the space $\widehat{\Phi}$, we must choose local coordinates which is equivalent to choosing a gauge. We choose a work in the global Coulomb coordinates i.e., $\partial_{i}A_{i}=0$. After substituting $A_{0}$, the Lagrangian may be written in terms of the orbit space variables $A_{i}\in \widehat{\Phi}$ (or $A_{i}$ is the so-called transverse potential)
\begin{eqnarray}
\mathcal{L}=\frac{1}{2}\int_{\mathbb{R}^{3}\times \mathbb{R}^{3}}\delta(x-y)\delta_{ij}\partial_{t}A_{i}(x)\partial_{t}A_{j}(y)\\\nonumber -\frac{1}{4}\int_{\mathbb{R}^{3}}F_{ij}F_{ij}.
\end{eqnarray}
Immediately I can read off the Riemannian metric that is induced on $\widehat{\Phi}$ by the kinetic energy 
\begin{eqnarray}
\mathcal{G}_{A_{i}(x)A_{j}(y)}=\delta(x-y)\delta_{ij}.
\end{eqnarray}
This is nothing but a flat metric. For our purpose, we need an additional piece, the measure on the space $\widehat{\Phi}$ that is obtainable through the ground state wave functional. Fortunately, one may exactly solve for the ground state wave functional (see \cite{wheeler1962geometrodynamics, hatfield2018quantum} for example) 
\begin{eqnarray}
\label{eq:U1}
\Psi[A]=N_{\hbar}e^{-2S[A]/\hbar}\\\nonumber =N_{\hbar}e^{-\frac{1}{2\hbar\pi^{2}}\int_{\mathbb{R}^{3}\times \mathbb{R}^{3}}\frac{(\nabla\times A(x))\cdot (\nabla\times A(y))}{|x-y|^{2}}}
\end{eqnarray}
to yield the necessary metric measure space for the $U(1)$ theory $(\widehat{\Phi},\mathcal{G},N_{\hbar}e^{-\frac{1}{2\hbar\pi^{2}}\int_{\mathbb{R}^{3}\times \mathbb{R}^{3}}\frac{(\nabla\times A(x))\cdot (\nabla\times A(y))}{|x-y|^{2}}})$.\\

\noindent A calculation for the associated Bakry-Emery curvature yields  
\begin{eqnarray}
\nonumber\text{Ricci}^{B.E}(\alpha[A],\alpha[A]):=\\\nonumber
\int\left((\mathcal{G}^{-1}_{\chi})^{A_{l}(z)A_{k}(y)}\mathcal{R}_{A_{l}(z)}~^{A_{i}(x^{''})}~_{A_{k}(y)}~^{A_{j}(x)}\right.\\\nonumber 
\left.\alpha_{A_{i}(x^{"})}[A]\alpha_{A_{j}(x)}[A]+\frac{2}{\hbar}(\mathcal{G}^{-1})^{A_{i}(x)A_{j}(x^{'})}(\mathcal{G}^{-1})^{A_{k}(y)A_{l}(y^{'})}\right.\\\nonumber 
\left.\frac{\mathfrak{D}}{\mathfrak{D}A_{i}(x)}\frac{\mathfrak{D}S}{\mathfrak{D}A_{k}(y)}\alpha_{A_{j}(x^{'})}[A]\alpha_{A_{l}(y^{'})}[A]\right)
\\\nonumber=\underbrace{0}_{flatness~of~the~orbit~space}\\\nonumber +\int_{x,x^{'},y,y^{'}}\frac{2}{\hbar}(\mathcal{G}^{-1})^{A_{i}(x)A_{j}(x^{'})}(\mathcal{G}^{-1})^{A_{k}(y)A_{l}(y^{'})}\\\nonumber \left.\frac{\mathfrak{D}}{\mathfrak{D}A_{i}(x)}\frac{\mathfrak{D}}{\mathfrak{D}A_{k}(y)}(\int_{w_{1},w_{2}}\frac{(\nabla\times A(w_{1}))\cdot (\nabla\times A(w_{2}))}{|w_{1}-w_{2}|^{2}})\right.\\\nonumber\left.\alpha_{A_{j}(x^{'})}[A]\alpha_{A_{l}(y^{'})}[A]\right)\\\nonumber 
\geq 0
\end{eqnarray}
which implies 
\begin{eqnarray}
\Delta E\geq 0.
\end{eqnarray}
This result depicts the well-known fact that there is no mass gap in $U(1)$ gauge theory i.e., photons can propagate freely thanks to the lack of confinement. Now we turn to the non-abelian gauge theories.\\
\section{Non-abelian Yang-Mills theory}
\noindent A gauge theory with a non-abelian compact semi-simple structure group differs from $U(1)$ gauge theory in a crucial way that involves the existence of a positive mass gap and possible confinement of the gauge bosons \cite{jaffe2006quantum}. Apart from these non-perturbative distinctions, it is well known that the non-abelian gauge theories enjoy asymptotic freedom \cite{gross1973ultraviolet}. Let us denote the Lie algebra valued connection $1-form$ of a principle bundle (pulled back by suitable section) by $A:=A^{P}_{\mu}e^{P}dx^{\mu}$, where $\{e^{P}\}$ constitute a basis for the associated Lie-algebra (see \cite{dewitt1978analysis} for a rigorous definition of a Yang-Mills theory on $\mathbb{R}^{1+3}$). The corresponding Yang-Mills curvature $2-$form reads $F:=F^{P}_{\mu\nu}e^{P}dx^{\mu}\wedge dx^{\nu}$, where the components in local coordinates are expressible as $F^{P}_{\mu\nu}:=\partial_{\mu}A^{P}_{\nu}-\partial_{\nu}A^{P}_{\mu}+[A_{\mu},A_{\nu}]^{P}$.  The Yang-Mills Lagrangian reads 
\begin{eqnarray}
\mathcal{L}=-\frac{1}{4}\int_{\mathbb{R}^{3}}\text{tr}(F\wedge ~^{*}F),
\end{eqnarray}
where $\text{tr}$ denotes a positive definite norm defined on the Lie algebra (From now on we will simply denote it by repeated gauge indices). Once again, $\Phi$, the space of connections lying in a suitable function space (e.g., Schwartz space) is not the true configuration space due to the gauge redundancy $A_{\mu}\sim g^{-1}A_{\mu}g+g\partial_{\mu}g^{-1}$, where $g$ is an element of the group of gauge transformations (let us denote it by $\mathfrak{G}$ and $\widehat{\mathfrak
{G}}$ be the group of reduced gauge transformation after modding out the center of $\mathfrak{G}$). The true configuration space of the Yang-Mills theory (also known as the orbit space) is then $\widehat{\Phi}:=\Phi/\widehat{\mathfrak
{G}}$. Similar to the $U(1)$ theory, in order to descend down to the orbit space $\widehat{\Phi}$, I need to eliminate $A^{P}_{0}$ by means of imposing the Gauss Law constraint ($\widehat{\nabla}^{2}A^{P}_{0}=\partial_{t}\partial_{i}A^{P}_{i}+[A_{i},\partial_{t}A_{i}]^{P},$ $\widehat{\nabla}^{2}$ is the gauge covariant Laplacian $\Delta_{A}$ defined below) and then choose a coordinate on the space $\widehat{\Phi}$. I choose a Coulomb coordinate around $A=0$ ($\partial_{i}A^{P}_{i}=0$) in which the coordinate invariant functional Lagrangian in terms of the orbit space coordinate reads (i.e., from now on the connection or the transverse potential is understood to be an element of the orbit space $\widehat{\Phi}$) 
\begin{eqnarray}
\mathcal{L}=\int_{\mathbb{R}^{3}\times \mathbb{R}^{3}}\frac{1}{2}\mathcal{G}[A]_{A^{P}_{i}(x)A^{Q}_{j}(x^{'})}\partial_{t}A^{P}_{i}(x)\partial_{t}A^{Q}_{j}(x^{'})\\\nonumber 
-\frac{1}{4}\int_{\mathbb{R}^{n}}F^{P}~_{ij}F^{P}~_{ij},
\end{eqnarray}
where the Riemannian metric $\mathcal{G}[A]_{A^{P}_{i}(x)A^{Q}_{j}(x^{'})}$ reads 
\begin{eqnarray}
\label{eq:metric}
\mathcal{G}[A]_{A^{P}_{i}(x)A^{Q}_{j}(x^{'})}=\delta_{ij}\delta_{PQ}\delta(x-x^{'})\\\nonumber+f^{PRV}A^{V}_{i}(x)\Delta^{-1}_{A}(x,x^{'})f^{RUQ}A^{U}_{j}(x^{'}),
\end{eqnarray}
where $f^{PQR}$ are the structure constants in a chosen Lie-algebra basis i.e., $[A_{i},A_{j}]^{P}=g^{2}_{YM}f^{PQR}A^{Q}_{i}A^{R}_{j}$. In particular, we absorb the factor $\sqrt{-1}$ within the structure constant $f$ and we denote the Yang-Mills coupling constant by $g_{YM}$. In addition $\Delta_{A}:=\widehat{\nabla}\cdot\widehat{\nabla}$ is the gauge covariant Laplacian ($\widehat{\nabla}_{i}:=\nabla_{i}+[A_{i},\cdot]$) (I want to mention that the metric expression (\ref{eq:metric}) is valid only in a small enough neighborhood of the flat connection $A=0$. In other words, we need multiple coordinate charts to cover the reduced configuration space. This is tied to the so-called \textit{Gribov ambiguity} of non-abelian gauge theory. This however does not cause any issue by the covariance of the formalism. In a generalized Coulomb chart around any oher connection $\overline{A}$, the metric would read $\mathcal{G}[A]_{A^{P}_{i}(x)A^{Q}_{j}(x^{'})}=\delta_{ij}\delta_{PQ}\delta(x-x^{'})+f^{PRV}(A-\overline{A})^{V}_{i}(x)\Delta^{-1}_{A}(x,x^{'})f^{RUQ}(A-\overline{A})^{U}_{j}(x^{'})$). The Riemannian nature of the metric follows from the compactness of the gauge group. As it turns out, the metric on $\widehat{\Phi}$ induced by the kinetic energy is curved and one may explicitly compute the Riemann curvature to yield 
\begin{eqnarray}
\mathcal{R}(W,Z,X,Y)=-2\langle[Y_{j},W_{j}],\Delta^{-1}_{A}[X_{i},Z_{i}]\rangle\\\nonumber -\langle[Z_{j},W_{j}],\Delta^{-1}_{A}[X_{i},Y_{i}]\rangle\nonumber+\langle[X_{j},W_{j}],\Delta^{-1}_{A}[Z_{i}, Y_{i}]\rangle
\end{eqnarray}
for $(X, Y, Z, W)\in T_{A}\widehat{\Phi}$ and the inner product $\langle\cdot,\cdot\rangle$ is the adjoint-invariant inner product on the Lie-algebra (I note that \cite{singer1981geometry,babelon1981riemannian,orland1} calculated the curvature and other geometric entities associated with the orbit space of the non-abelian Yang-Mills theory as well). The sectional curvature that determines the Riemann curvature completely is computed to be \begin{eqnarray}
\mathcal{K}_{X,Y}=3\langle[X,Y],\Delta^{-1}_{\widehat{A}}[X,Y]\rangle.
\end{eqnarray}
which manifestly enjoys a positive definite property. Unfortunately, the Ricci curvature as a trace of the Riemann curvature is not well defined as a consequence of the Riemann curvature not being a trace-class operator. This feature is usually present in infinite dimensions (See \cite{freed1989basic} for example). The formal expression of the Ricci curvature is as follows 
\begin{eqnarray}
\text{Ricci}(X,Y)=\\\nonumber3f^{VPR}X^{R}_{i}(x)\text{tr}\Delta^{-1}_{A}(x,x^{'})f^{VPU}Y^{U}_{i}(x^{'}).
\end{eqnarray}
However, $\text{tr}\Delta^{-1}_{A}(x,x^{'})$ is divergent. This follows from the simple observation that at $A=0$, one has $\Delta^{-1}(x,x^{'})=-\frac{1}{4\pi}|x-x^{'}|^{-1}$ that blows up as $x\to x^{'}$. Therefore, I need to regularize the Ricci curvature to make sense of it. Fortunately, Ricci curvature appears in the mass gap expression in a regularized fashion (point spit or otherwise). This is a consequence of the fact that Ricci curvature and the Laplacian are of the same type as operators on a manifold. One may compute the regularized Ricci curvature at the flat connection $A=0$ to yield 
\begin{eqnarray}
\text{Ricci}_{\chi}[A=0](\alpha,\alpha)=\frac{3\chi C_{2}g^{2}_{YM}}{2\pi^{3}}\\\nonumber \int_{x,x^{'}}\alpha^{P}(x)\alpha^{P}(x^{'})d^{3}xd^{3}x^{'},
\end{eqnarray}
where $\chi$ is the regularization scale and $C_{2}$ is the Casimir invariant of the adjoint representation of the Lie algebra. The choice of a regularization scale implies the breaking of scale invariance upon quantization that is present in classical theory. In order for the regularized Ricci curvature to be independent on $\chi$, $g^{2}_{YM}$ must be a function of $\chi$ i.e., $g_{YM}$ becomes a \textit{running} coupling constant $g^{2}_{YM}=g^{2}_{YM}(\chi)$. Since, in the renormalizable procedure one observes that $g^{2}_{YM}$ becomes a function of the energy scale $\chi$, a more appropriate expression for the regularized curvature would be as follows 
\begin{eqnarray}
\text{Ricci}_{\chi}[A=0](\alpha,\alpha)=\frac{3 C_{2}\chi g^{2}_{YM}(\chi)}{2\pi^{3}}\\\nonumber \int_{x,x^{'}}\alpha^{P}(x)\alpha^{P}(x^{'})d^{3}xd^{3}x^{'},
\end{eqnarray}
and one should expect in the regularization limit i.e., $\chi\to\infty$ the following holds 
\begin{eqnarray}
 \lim_{\chi\to\infty}g^{2}_{YM}(\chi)\chi=m_{0}   
\end{eqnarray}
In $3+1$ dimensions, the introduction of an energy scale $m_{0}$ through renormalization seems inevitable purely on the dimensional ground. The classical action $\int_{\mathbb{R}^{1,3}}\langle F,F\rangle dt d^{3}x$ has the dimension of $\hbar$. Therefore the connection $A$ has the dimension of $\frac{\hbar^{\frac{1}{2}}}{L}$ and $g^{2}_{YM}$ has the dimension of $\frac{1}{\hbar}$. Now the dimension of $\frac{\mathfrak{R}ic_{\chi}(\alpha,\alpha)}{\int_{x,x^{'}}\alpha(x)\alpha(x^{'})}$ is $\frac{1}{\hbar L}$ since $\chi$ has dimension $\frac{1}{L}$. Therefore the entity $\Delta$ in the gap theorem has dimension $\frac{1}{\hbar L}$ yielding the dimension of $\Delta E$ to be $\frac{\hbar}{L}$ which is the correct dimension of energy. Introduction of the energy scale $m_{0}$ essentially breaks the conformal invariance of the Yang-Mills theory at the quantum level. These ideas are to be understood from a rigorous non-perturbative renormalization group flow perspective. Nevertheless, `dimensional transmutation', the phenomenon of introducing an energy scale is well understood by particle physicists (see \cite{coleman1973radiative}). In summary, regularized Ricci curvature exhibits a positive definiteness property at the flat connection $A=0$. One may utilize a heat kernel argument to show that the regularized Ricci curvature is strictly positive definite away from the flat connection or more precisely
\begin{eqnarray}
\text{Ricci}_{\chi}[A\neq 0](\alpha,\alpha)>\text{Ricci}_{\chi}[A=0](\alpha,\alpha)\\\nonumber=\frac{3m_{0} C_{2}}{2\pi^{3}} \int_{x,x^{'}}\alpha^{P}(x)\alpha^{P}(x^{'})d^{3}xd^{3}x^{'}.
\end{eqnarray}
We do not present the calculations here since it is presented elsewhere \cite{puskar2023geometric}.
Therefore the mass gap $\Delta E$ verifies
\begin{eqnarray}
\Delta E\geq \frac{3\hbar^{2}m_{0}C_{2}}{4\pi^{3}}+I,
\end{eqnarray}
where $I$ verifies 
\begin{eqnarray}
I=\frac{2}{\hbar}\int_{\widehat{\Phi}}\left(\int_{(\mathbb{R}^{3})^{4}}(\mathcal{G}^{-1})^{A^{p}_{i}(x)A^{q}_{j}(x^{'})}\right.\\\nonumber\left. (\mathcal{G}^{-1})^{A^{r}_{k}(y)A^{s}_{l}(y^{'})} \frac{\mathfrak{D}}{\mathfrak{D}A^{p}_{i}(x)}\frac{\mathfrak{D}S}{\mathfrak{D}A^{r}_{k}(y)} \alpha_{A^{q}_{j}(x^{'})}[A]\right.\\\nonumber\left.\alpha_{A^{s}_{l}(y^{'})}[A]\right)|N_{\hbar}|^{2}e^{-2S[A]/\hbar}/\\\nonumber \int_{\widehat{\Phi}}\left(\int_{\mathbb{R}^{3}\times \mathbb{R}^{3}}\mathcal{G}^{A^{p}_{i}(x)A^{q}_{j}(y)}\alpha_{A^{p}_{i}(x)}[A]\alpha_{A^{q}_{j}(y)}[A]\right)\\\nonumber |N_{\hbar}|^{2}e^{-2S[A]/\hbar}
\end{eqnarray}
The fundamental difficulty now lies in estimating $I$ and proving its non-negative definiteness property. At the level of perturbation calculations, the lowest order term of the Hessian functional in the coupling constant is the same as that of the Maxwell theory and therefore enjoys a non-negative definite property. However, notice that the expression of $I$ is fully non-perturbative as it involves integration over the manifold $\widehat{\Phi}$. We present a physical picture to argue why our expectation is that $\Delta E$ is strictly positive.

The $S$ functional appearing in the gap estimate encodes the processes arising from the self-interacting Yang-Mills potential. Potential admitting flat directions (lower-dimensional varieties of the reduced configuration space $\widehat{\Phi}$) may give rise to mass-less modes (one important example is the excitation of the goldstone modes in the framework of spontaneous symmetry breaking). The Yang-Mills potential exhibits flat directions whenever the commutator $[A, A]$ vanishes. Since such flat directions do not cost any energy, the excitation of mass-less modes becomes plausible. Therefore the Hessian contribution above can vanish on such lower dimensional spaces of the orbit space $\widehat{\Phi}$. In fact, the Hessian of a gauge invariant entity such as $S$ is bound to have a finite index (number of negative eigenvalues) at each connection since the topology of $\widehat{\Phi}$ is non-trivial. However, since the number is finite, they supposedly integrate out to zero in the measure-theoretic sense. Making a mathematically rigorous sense of such an argument is currently lacking.  However, notice that in the expression for $\Delta E$, the uniform strictly positive lower bound $\frac{3\hbar^{2}m_{0}C_{2}}{4\pi^{3}}$ persists even if the potential contribution $I$ vanishes. In other words, due to the presence of a non-trivial (positive) regularized Ricci curvature of the configuration space $\widehat{\Phi}$, one requires energy for any excitation leading to a non-trivial mass gap in the spectrum. There is a study in a pure quantum mechanical setting that proves the existence of a spectral gap in a system with potential admitting flat directions \cite{simon1983some} generalization of which to the field-theoretic settings is not known. It would seem to solve the quantum Yang-Mills theory if one were to obtain the $S$ functional i.e., prove its existence, growth at infinity (of $\widehat{\Phi}$) among other desirable properties. There are proposed techniques to obtain the $S$ functional in a gauge invariant non-perturbative way based on a microlocal type approach \cite{moncrief2,marini} and also through different approaches \cite{nair2012quantum}. It would be interesting to pursue a possible connection between my geometric approach and more direct approaches such as solving the Dyson-Schwinger equations \cite{frasca1,frasca2}. Another interesting question would be whether the renormalization group flow of the quantum Yang-Mills theory can be cast into a Ricci flow in the moduli space of metrics $\mathcal{G}$ indexed by the coupling constant $g_{YM}$. If so, then the analytic methods of Ricci flow can be utilized to assert that the positive sign of the Ricci curvature (regularized) should indicate the existence of a gapped spectrum and a trivial theory at high energy leading to asymptotic freedom.\\
One may compute the Hessian term $I$ for two extreme cases: at the flat connection $A=0$ (also high energy or $g_{YM}\to 0$ limit) and low energy or large length scale limit. In both of these limits, one obtains a non-negative contribution. The formal wave functional at a high energy limit $g_{YM}\to 0$ reads \cite{hatfield1984first,hatfield2018quantum}
\begin{eqnarray}
N_{\hbar}e^{-\frac{S[A]}{\hbar}}=N_{\hbar}e^{-\frac{1}{2\hbar\pi^{2}}\int_{\mathbb{R}^{3}\times \mathbb{R}^{3}}\frac{(\nabla\times A(x))\cdot (\nabla\times A(y))}{|x-y|^{2}}+O(A^{3})}
\end{eqnarray}
and therefore the Hessian at $A=0$ is non-negative definitive. At the low energy (or strong coupling) limit, it is conjectured \cite{greensite1979calculation} that the wave functional should be of the \textit{magnetic} type (at large scale chromoelectric field is suppressed in non-abelian theory)
\begin{eqnarray}
N_{\hbar}e^{-\frac{S[A]}{\hbar}}\sim N_{\hbar}e^{-\frac{1}{2\mu\hbar}\int_{\mathbb{R}^{3}}F^{p}_{ij}F^{p}_{ij}}
\end{eqnarray}
for some mass dimension $\mu>0$. Once again, the Hessian term contributes by a non-negative continuous factor on non-measure-zero sets (roughly $\sim O(k^{2})$). Technically, the hessian of the $S$ functional in this limit has a finite index at each connection (see \cite{taubes1983stability} for the estimate of the index of Hessian of Yang-Mills action functional at critical points). However, a natural expectation is that the finiteness of the index essentially does not affect in the sense of zero measure (note that the Hessian of $S$ functional is integrated over the entire orbit space and therefore finitely many directions are supposed to yield measure zero upon integration). The remaining task would be to interpolate between these two length scales. Due to Lorentz covariance, one would expect that the contribution of the Hessian term $I$ is not independent of the kinetic part and it should produce a contribution of the type $\sqrt{m^{2}_{0}+k^{2}}$ upon summation to all orders. A substantial amount of work is done in both $2+1$ and $3+1$ dimensional cases (see \cite{nair2012quantum} for $2+1$ dimensions and \cite{mansfield1999yang} for $3+1$ dimensions). However, the complete construction of the ground state Yang-Mills wave function remains far beyond the current quantum field theory. One would like to understand if there is a clear argument based on re-normalization group flow. Nevertheless, from a physical perspective, it seems that the potential contribution that is manifested through the Hessian term $I$ should produce a positive factor due to Lorentz invariance. This remains to be studied in detail.

Another interesting point worth noting is the behavior of the Ricci curvature (and hence the mass gap) at a large $N$ limit in the case when the structure group is $SU(N)$. More precisely, keeping the t'Hooft coupling $\lambda=g^{2}_{YM}N$ fixed, the limit $N\to\infty$ is defined to be the large $N$ limit \cite{t1993planar}. Remarkably notice that the curvature contribution of the mass-gap $\Delta E_{curvature}=\frac{3\chi C_{2}g^{2}_{YM}}{2\pi^{3}}$ remains unchanged since $C_{2}(SU(N))=N$ and therefore $\Delta E_{curvature}=\frac{3\chi \lambda}{2\pi^{3}}$ remains unchanged as long as $\lambda$ is fixed. The strict $N\to\infty$ limit is essentially a free theory in the sense that all the correlation functions of single trace, gauge invariant operators factorize (maps onto a free string theory\cite{hatfield2018quantum,t1993planar}). Nevertheless, in the large $N$ limit, the theory exhibits a mass gap (in fact the strict lower bound does not depend on $N$ as long as the t'Hooft coupling is fixed as seen from the explicit expression). Therefore in the strict large $N$ limit, we have a tower of massive free particles. The lowest mass is essentially expressed in terms of the lowest eigenvalue of the regularized Ricci curvature of the orbit space. It would be interesting to make sense of these rather heuristic arguments in a mathematically rigorous way. Lattice gauge theory calculations are suggestive of a strictly positive mass gap and therefore color confinement in $2+1$ and $3+1$ dimensions  \cite{athenodorou2020glueball,lucini2010glueball}.

\noindent \section{Scalar Electrodynamics}
\noindent The last model that we should consider is mass-less scalar electrodynamics on $(\mathbb{R}^{1+3},\eta)$.  The Lagrangian reads 
\begin{eqnarray}
\mathcal{L}=\int_{\mathbb{R}^{3}}\left(-\frac{1}{4}F\wedge F-\eta^{\mu\nu}(D_{\mu}\varphi)^{\dag}(D_{\nu}\varphi)\nonumber+U[(\varphi^{\dag}\varphi)^{2}]\right),
\end{eqnarray}
where $F$ is the curvature of a $U(1)$ bundle, $\varphi:=\varphi^{1}+\sqrt{-1}\varphi^{2}$, with $\varphi^{1,2}$ real, a complex scalar field. In terms of the connection $A$ ($F=dA$), the action of the gauge-covariant derivative $D$ on a section $\varphi$ reads 
\begin{eqnarray}
D_{\mu}\varphi:=\partial_{\mu}\varphi-\sqrt{-1}eA_{\mu}\varphi,\\\nonumber (D_{\mu}\varphi)^{\dag}:=\partial_{\mu}\varphi^{\dag}+\sqrt{-1}eA_{\mu}\varphi^{\dag}.
\end{eqnarray}
Here $e$ is the charge in the units $\hbar=1=c$. The group of gauge transformation $\mathfrak{G}$ acts as $A_{\mu}\mapsto A_{\mu}+\partial_{\mu}\Lambda$, $\varphi\mapsto e^{ie\Lambda}\varphi$ for a smooth function $\Lambda$ vanishing at $\infty$. Let the space of connections and space of complex scalar fields be denoted by $\mathcal{A}$ and $\mathcal{S}$, respectively. The true configuration space is essentially $\widehat{\Phi}:=\frac{\mathcal{A}\times \mathcal{S}}{\mathfrak{G}}$. We can obtain a Riemannian metric on $\widehat{\Phi}$ through the kinetic energy. \cite{moncrief} has obtained this metric that reads in a global Coulomb coordinates $(A|\partial\cdot A=0,\varphi)$
\begin{eqnarray}
\mathcal{G}_{A(x)A(y)}=\delta(x-y),\\
\mathcal{G}_{\varphi^{a}(x),\varphi^{b}(y)}=2\delta_{ab}\delta(x-y)\\\nonumber +4e^{2}\epsilon_{ac}\varphi^{c}(x)\Delta^{-1}_{\varphi}(x,y)\epsilon_{bd}\varphi^{d}(y),
\end{eqnarray}
where $\Delta_{\varphi}:=\Delta-2e^{2}\varphi^{\dag}\varphi$ and $\epsilon$ is the usual $2-$dimensional alternating symbol. Notice that the metric of the orbit space $\widehat{\Phi}$ is block-diagonal (reducible). The corresponding $U(1)$ sector is flat leading to the zero curvature contribution to the mass of $U(1)$ boson. On the other hand, the Ricci curvature associated with the $\varphi$ sector is non-vanishing. A simple calculation yields the following expression for the formally positive definite Ricci curvature 
\begin{eqnarray}
\text{Ricci}[\varphi=0](\alpha,\alpha)\\\nonumber=-6e^{2}\int_{x,y}\alpha_{a}(x)(\text{tr}\Delta^{-1}(x,y))\alpha^{a}(y),\\
\alpha\in T_{A,\varphi}\widehat{\Phi}
\end{eqnarray}
which after regularization yields a finite positive result. Using a heat kernel argument, one can show that the Ricci curvature is modified by a positive factor away from $\varphi=0$. As a consequence of the gap theorem, the scalar field acquires mass upon quantization while photons remain massless. The regularization and renormalization

\noindent This work is supported by Harvard CMSA and Mathematics department.

\end{document}